\documentclass[proceedings,a4paper]{JHEP3}

\def\br{\begin{eqnarray}}
\def\er{\end{eqnarray}}
\def\be{\begin{equation}}
\def\ee{\end{equation}}
\def\({\left(}
\def\){\right)}

\def\pa{\partial}
\def\u2{\mid u\mid^2}
\def\h{{1\over 2}}
\def\l{\lambda}

\def\rlx{\relax\leavevmode}
\def\IZ{\rlx\hbox{\sf Z\kern-.4em Z}}
\newcommand{\sbr}[2]{\left\lbrack\,{#1}\, ,\,{#2}\,\right\rbrack}
\def\tr{\mathop{\rm tr}}

\def\rf#1{(\ref{eq:#1})}
\def\lab#1{\label{eq:#1}}
\def\ca{{\cal A}}

\def\cf{{\cal F}}
\def\cg{{\cal G}}

\def\cj{{\cal J}}

\def\nonu{\nonumber}
\def\o{\over}
\def\PRL#1#2#3{{\sl Phys. Rev. Lett.} {\bf#1} (#2) #3}
\def\NPB#1#2#3{{\sl Nucl. Phys.} {\bf B#1} (#2) #3}

\def\PRD#1#2#3{{\sl Phys. Rev.} {\bf D#1} (#2) #3}

\def\PLB#1#2#3{{\sl Phys. Lett.} {\bf #1B} (#2) #3}
\def\JMP#1#2#3{{\sl J. Math. Phys.} {\bf #1} (#2) #3}
\def\vecgrad{{\vec{\nabla}}}
\def\u2{\mid u\mid^2}
\def\z2{\mid z\mid^2}

\conference{Workshop on integrable theories, solitons and duality.}
\title{On the connections between  
Skyrme  and Yang Mills theories}
\author{\speaker{Joaqu\'{\i}n S\'anchez Guill\'en}\\
Facultad de F\'\i sica USC,\\
15706 Santiago de Compostela, Spain \\
E-mail \email{joaquin@fpaxp1.usc.es}  }
\author{L.A. Ferreira \\           
Instituto de F\'\i sica Te\'orica - IFT/UNESP\\
Rua Pamplona 145\\
01405-900  S\~ao Paulo-SP, Brazil \\
E-mail \email{laf@ift.unesp.br}  }
\abstract{Skyrme theories on $S^3$  and $S^2$, are analyzed using the 
generalized zero curvature in any dimensions.
 In the first case, new symmetries and integrable
sectors, including the $\mid B\mid =1$ skyrmions,  are unraveled.
In $S^2$ the relation to QCD suggested by Faddeev is discussed.}

\begin{document}

\section{Introduction}

Skyrme   theory, based on   chiral fields with an stabilizing  quartic
derivative term \cite{Sky} at the classical level,
 was  an alternative to the standard field theory approach in ideas and
methods.
The  theory was shown later to correspond to the non-abelian  gauge theory
with expansions in
number of colours (soliton aspects) \cite{Witten} and momentum
(chiral aspects). 
Faddeev conjectured a more direct connection to pure QCD, restricting the
Skyrme chiral fields to the coset
$SU(2)/U(1)$ \cite {Fad}.
Non-perturbative  progress  generally used numerical methods  both for
ordinary Skyrme \cite{Houg} as well as for
Faddeev  $\sigma$-model formulation \cite {Fade}, which has been also
investigated on the lattice  \cite {vW}.
A  generalization of  the zero-curvature methods of two dimensional field
theory to higher  dimensions \cite{afg} offered a new
possibility  for analytical progress,
 in a scheme which uses gauge techniques and fields as auxiliary
connections to study
non-linear  systems. A zero curvature representation for Skyrme-Faddeev
theory was given in \cite{afz} among other examples of models
 defined on the sphere $S^2$, 
and discussed in \cite{Last} in connection with QCD. The
integrable sector of the  $S^3$ Skyrme theory corresponding to $\mid
B\mid =1$, was found in \cite {last}.

Here  we review and  explore further  both  theories with that integrablity
method, which is very briefly summarized in next section 2,
devoted to the ordinary Skyrme. The gauge ambiguities of the method are
exhibited and the choice of the hedgehog
Ansatz, found by direct computation in \cite {last}, is explained. In section
3 we work out in detail the Skyrme Faddeev case,
and we discuss how
the gauge formalism of the method clarifies the observations \cite{BW} and
conjectures of a
connection between Skyrme-Faddeev theory and the long distance limit of the non
abelian gauge theory.

\section{The Skyrme model}
\label{app: Skyrme}
The Lagrangian density for the Skyrme model can be written as:
\be
L = - {f_\pi^2 \over 4}
\tr \left ( U^\dagger \pa_\mu U U^\dagger \pa^\mu U \right ) +
{1 \over 32 e^2}
\tr \left [ U^\dagger\pa_\mu U,U^\dagger\pa_\nu U \right ]^2
\ee
where $f_\pi$ and $e$ are phenomenological constants, and $U$ is an unitary
matrix representation of a compact Lie group $G$. The cases of physical
interest correspond to $G$ being $SU(2)$ or $SU(3)$. In terms of the
Lie algebra valued field
$A_\mu = U^\dagger \pa_\mu U= A_\mu^{i} T_i$ we can write the Lagrangian as:
\be
L = - {f_\pi^2 \over 4}
\tr \left ( A_\mu A^\mu \right ) +
{1 \over 32 e^2}
\tr \left \{ [ A_\mu , A_\nu ] [A^\mu,A^\nu] \right \}
\ee

The equations of motion which can be derived from this Lagrangian are:
\be
\label{em}
\pa_\mu \left ( A^\mu - {\epsilon} [A_\nu,[A^\mu,A^\nu]]
\right ) = 0
\lab{skyrmeeqom}
\ee
where $\epsilon=1/4f_\pi^{2} e^2$.

Let us define the auxiliary field
\be
\tilde{J}^\mu = A^\mu - {\epsilon} [A_\nu,[A^\mu,A^\nu]]
\lab{hdef}
\ee
The equations of motion can then be written in the form $\pa_\mu
\tilde{J}^\mu =0$.
The space components of the second term in \rf{hdef} can be normalized to
the degree of the map $S^3\rightarrow S^3$, which gives a
topological meaning to 
the baryon  number of the solution.
Squaring it one gets a lower bound for the energy functional in a
given charge sector, which unfortunately 
can only be saturated
in 3 spatial dimensions  by $A=0$. In addition, the bound does not lead
to a lower degree 
 equation of the BPS type
\cite{Novikov}. In fact, the only known exact solution is the original
$B=1$  {\it hedgehog}  Ansatz for the static Skyrme field:
\be
\label{hedge}
U(\vec{x}) = \exp\left(i \hat{r}\cdot \vec{\tau} f(r)\right)
\ee
where $r=|\vec{x}|$ and $\hat{r} = {\vec{x}\over r}$,
$\vec{\tau}$ are the Pauli matrices, and $f(r)$ is the profile
function. With this unique maximally symmetric Ansatz, it is well known
\cite{brown} that the equations of
motion \rf{skyrmeeqom} reduce to an
ordinary differential equation in $f$, which   has then to be handled
numerically (but being an ODE it is an existence proof).
One way to progress  in  the  analytical understanding of the Skyrme
model, is to study its equations of motion with the
geometric approach of ref. \cite{afg}  as we now
explain.

The method is  a generalization, for a $(d+1)$-dimensional
space-time,  of the well known two
dimensional Lax-Zakharov-Shabat zero curvature condition. The construction
involves a flat connection on the space of $(d-1)$-loops
(closed $(d-1)$-dimensional hypersurfaces) which is built from a
$1$-form $\ca$ and a $d$-form $B$ on space-time. It is possible to find local
sufficient conditions on the latter for the loop space connection to be
flat. Those conditions involve a non-semisimple Poincar\'e type algebra which
decomposes into a Lie algebra $\cg$ and an invariant abelian subalgebra $P$
transforming under some representation $R$ of $\cg$. The local zero curvature
conditions are given by
\br
\cf_{\mu\nu} = \pa_{\mu} \ca_{\nu} - \pa_{\nu} \ca_{\mu}
+ \sbr{\ca_{\mu}}{\ca_{\nu}}
= 0 \; ; \qquad
D_\mu \tilde{B}^\mu = \pa_\mu \tilde{B}^\mu +
\sbr{\ca_\mu}{\tilde{B}^\mu}=0
\lab{zccond}
\er
where $\tilde{B}^\mu$ is the dual of the $d$-form refered above.

In our $d=3$ Skyrme case, since $A_{\mu}= U^{\dagger} \pa_{\mu} U$ is
flat by construction, 
it is natural to start with  $\ca_{\mu} \equiv A_{\mu}$. 
Take  $\tilde{B}_\mu
\equiv \tilde{J}_\mu^a P_a$, where we have written $\tilde{J}_\mu =
\tilde{J}_\mu^a T_a$, with $T_a$'s being the generators of $\cg$, and $P_a$'s
transforming under the adjoint
representation of $\cg$, i.e. $\sbr{T_a}{T_b} = i f_{ab}^c T_c$,
$\sbr{T_a}{P_b}= i f_{ab}^c P_c$, and $\sbr{P_a}{P_b} = 0$. Notice
that the Jacobi identities require that
$[A_{\mu},\tilde{J}^\mu]=0$. Then 
it is clear that \rf{skyrmeeqom} is equivalent to \rf{zccond}. We have then
 expressed  Skyrme   equations as  local zero curvature conditions of
 \cite{afg}. 
\subsection{Constraints.The most simple case}
Notice that with $\ca_{\mu}= U^{\dagger} \pa_{\mu} U$, the quantities  
$J_{\mu} \equiv U \tilde{B}_\mu U^{\dagger}$ are  conserved
currents as a consequence of \rf{zccond}. Together with ${\tilde{J}^\mu}$
themselves, those are the Noether currents associated to the $G\otimes G$
global symmetry of the Skyrme model. If the equivalence holds only for 
$\tilde{B}_\mu$ being in the
adjoint representation, we have just reexpressed the
equations of motion 
with the geometric gauge formalism, while {\it if it holds for any
representation},
we can  discover  hidden symmetries, as in the $2d$ case of Sine
Gordon and
Toda theories. In the Skyrme model that can be implemented and the
formulation \rf{zccond} can be used  
to construct an infinite
number of conserved currents for some sectors of the Skyrme
theory. However, the sectors one gets depend crucially on the choice
(gauge) of the zero curvature potentials. In \cite{last} it was
constructed an integrable sector containing the charge $\pm 1$
skyrmions. Here we discuss the integrable sector of Skyrme theory
obtained from the   
choice above of $A$ and $B$. One can just follow the case of the chiral model
 \cite{gianzo} and  introduce,
for any integer spin $j$ representation of $SU(2)$, the
operator\footnote{Notice that the normalization of the coefficients are due to
the fact that $(- P^{(1)}_{+1})$, $(P^{(1)}_{0}/\sqrt{2})$ and $P^{(1)}_{-1}$
constitute the basis of the adjoint of $SU(2)$ that transforms exactly as
$T_{+}$, $T_3$ and $T_{-}$ respectively.}
\be
\tilde{B}_\mu^{(j)} = - \tilde{J}_\mu^{+} P^{(j)}_{+1}  +
{1\over\sqrt{j(j+1)}} \tilde{J}_\mu^{0} P^{(j)}_{0} +
\tilde{J}_\mu^{-} P^{(j)}_{-1}
\lab{bjdef}
\ee
where we have denoted the quantity \rf{hdef} as
$$
\tilde{J}_\mu = \tilde{J}_\mu^{+} T_+ +
\tilde{J}_\mu^{0} T_{3} + \tilde{J}_\mu^{-} T_-
$$
and where $T_{3,\pm}$ are the usual basis for the angular momentum algebra and
$P^{(j)}_m$ transform under the spin $j$ representation of $SU(2)$, i.e.
$\sbr{T_3}{T_{\pm}}= \pm T_{\pm}$, $\sbr{T_{+}}{T_{-}}= 2 T_3$,
$\sbr{T_3}{P^{(j)}_m}= m P^{(j)}_m$,
$\sbr{T_{\pm}}{P^{(j)}_m}= \sqrt{j(j+1)-m(m\pm 1)} \, P^{(j)}_{m\pm 1}$, and
$\sbr{P^{(j)}_m}{P^{(j)}_n} = 0$.

Denoting $A_{\mu} = A_{\mu}^+ T_{+} + A_{\mu}^0 T_3 + A_{\mu}^- T_{-}$, and
using the fact that $[A_\mu,\tilde{J}^\mu]=0$  we get that
$A^{\mu ,i} \tilde{J}_\mu^{j} - A^{\mu ,j} \tilde{J}_\mu^{i} = 0$, for
$i,j=0,\pm$. Consequently, for the spin $1$ representation we get
$\sbr{A^{\mu}}{\tilde{B}_\mu^{(1)}}=0$. However, for $j>1$ we get that
$\sbr{A^{\mu}}{\tilde{B}_\mu^{(j)}}=0$ if and only if
\be
A^{\mu ,+} \tilde{J}_\mu^{+}=A^{\mu ,-}\tilde{J}_\mu^{-}=0
\lab{constraint}
\ee

The conclusion we then reach is that if we substitute the operator \rf{bjdef}
into \rf{zccond}  with $\ca_{\mu} \equiv A_{\mu} = U^{\dagger}
\pa_{\mu} U$,  we get, for $j=1$, just the equations of motion for the
Skyrme model, namely 
$\pa^{\mu} \tilde{J}_\mu =0$. However, if we impose the constraints
\rf{constraint} we can get the same equations but with the zero
curvature potential ${\tilde B}$ being in any integer spin $j$
representation. That implies that the submodel of the Skyrme theory defined by
the equations
\be
\pa^{\mu} \tilde{J}_\mu =0 \; ; \qquad \qquad
A^{\mu ,+} \tilde{J}_\mu^{+}=A^{\mu,-}\tilde{J}_\mu^{-}=0
\ee
possesses an infinite number of conserved currents given by
\be
J_\mu^{(j)} = U\tilde{B}_\mu^{(j)}U^\dagger \equiv
\sum_{m=-j}^j J_\mu^{(j),m} \, P^{(j)}_m  \; ; \qquad \qquad
\mbox{\rm for any positive integer $j$}
\ee

\subsection {The sector of the skyrmion solution}
The restriction
\be
\label{restric}
A_\mu^\pm \tilde{J}^\mu_\pm = 0
\ee
is  highly
non-trivial, and it is not clear  whether the reduced model has any
solutions at all.\footnote{The constraint is analogous to the chiral model
\cite{gianzo}, where it is not difficult to
obtain solutions, although subject to the scaling instabilities.} For
the only  known Ansatz (\ref{hedge}),
it turns out that the constrained equations (\ref{restric}) in the
static case, restricts the profile function $f(r)$ severely. One finds that the
conditions (\ref{restric}) are solved by 
\be
\label{restsol}
f_R(r) = 2 {\rm ArcCotan}\, (c\,r)
\ee
where $c$ is a constant representing the (inverse) size parameter of the
extended solution. The configuration (\ref{restsol}) does not solve
the static equations of motion $\pa_k \tilde{J}_k=0$. However, it
approximates the solution for an interval of the radial variable $r$
which is of physical interest. Plugging (\ref{restsol}) into the
equations of motion, one gets a polynomial in $r$ of order
four. Solving 
it  implicitly for $ c$, for the physical values of the couplings, one finds
that there exist admissible solutions for values of
$r$ up to half a Fermi
{\it strongly peaked}  around  a very reasonable value of  $c$,
between $2$ and $3$ $Fm^{-1}$ for  $f_\pi$
in the typical range of 60 to 120 $GeV$. The minimum of the energy for these
lower and upper values  is 
$1$ and $2$ $GeV$ respectively, again as expected.
So, for practical purposes, we conclude that the restricted solution
(\ref{restsol})  is in fact a good approximation  for
values of $r$ of the order of the light particle sizes  and for the
physical values of the size parameter $c$.

It is also interesting that this simplified Ansatz was used in \cite{Rafael}
to argue the absence of stable
solutions in the Susy $CP^1$, although the authors warn for the
possibility that it might not be a
solution, as we see here in the related Skyrme case.

We have seen how the geometric method \cite{afg} works in the
construction of integrable submodels of the Skyrme theory. The great
problem is to find the gauge choice for the zero curvature potentials that
produce constraints compatible with the equations of motion. 
The physical solution of the Skyrmion turns out no to  be in the
 simple gauge chosen above, namely by starting directly with the
adjoint representation as in the case of the chiral model
\cite{gianzo}.

The correct choice of gauge to get the charge $\pm 1$ skyrmions inside
the integrable sector was presented in \cite{last}. One has to write
the group element as 
\be
U = W^{\dagger} \; e^{-i \zeta \tau_3}\; W
\lab{nicedecomp}
\ee
where $\tau_3$ is the diagonal Pauli matrix and 
\br 
W\equiv  \frac{1}{\sqrt{1+\u2}}\; \( 
\begin{array}{cc}
1 & i u\\
i u^* & 1 
\end{array}\) 
\lab{wdef}
\er
with $\zeta$ being a real scalar field, and $u$ a complex one. 
Then the zero curvature potentials are taken to be 
\br
\ca_{\mu} &\equiv & - \pa_{\mu} W\; W^{\dagger} 
= {1\o{ 1+\mid u \mid^2 }} \(  -i \pa_{\mu} u \, \tau_{+}
-i \pa_{\mu} u^* \, \tau_{-} +  
\h \; \( u \pa_{\mu} u^* - u^* \pa_{\mu} u \) \, \tau_3 \)
\lab{asky}\\
{\tilde B}_{\mu}&\equiv & -i R_{\mu} \tau_3 + \frac{2 \sin \zeta}{1+ \u2}\; \( 
e^{i\zeta}\; S_{\mu} \; \tau_{+} -
e^{-i\zeta}\; S_{\mu}^* \; \tau_{-} \) 
\lab{bsky}
\er
where
\br
R_{\mu}&\equiv& \pa_{\mu} \zeta - 8 \l \; \frac{\sin^2 \zeta}{\(1 + \u2\)^2}
\(  N_{\mu} +  N_{\mu}^*\)\nonu\\
S_{\mu}&\equiv& \pa_{\mu}u + 4 \l \; \( M_{\mu} - \frac{2\sin^2
\zeta}{\(1+\u2\)^2}  \; K_{\mu}\)
\lab{rsdef}
\er
and 
\br
K_{\mu} &\equiv& \( \pa^{\nu} u \pa_{\nu} u^*\) \pa_{\mu} u - \(\pa_{\nu}u\)^2
\pa_{\mu} u^*\nonu\\ 
M_{\mu} &\equiv& \( \pa^{\nu} u \pa_{\nu} \zeta\) \pa_{\mu} \zeta 
- \(\pa_{\nu}\zeta\)^2
\pa_{\mu} u\nonu\\ 
N_{\mu} &\equiv& \( \pa^{\nu} u \pa_{\nu} u^*\) \pa_{\mu} \zeta 
- \(\pa_{\nu}\zeta \pa^{\nu} u\)
\pa_{\mu} u^*
\lab{kmndef}
\er
One can check that the conditions \rf{zccond} with the potentials
\rf{asky} and \rf{bsky} are equivalent to the equations of motion 
\rf{skyrmeeqom}.

By extending the potential \rf{bsky} to any integer spin $j$
representation, in a manner similar to the one we did in \rf{bjdef},
one gets highly non-trivial constraints. However, in the static case
those constraints reduce to the conditions 
\be
\vecgrad u \cdot \vecgrad u =0 \; ; \qquad \qquad 
\vecgrad u \cdot \vecgrad \zeta =0
\lab{staticconst}
\ee
They are easily solved by the time independent configurations 
\be
\zeta = \zeta \( r\) \qquad u = u \( z \) \qquad u^* = u^* \( z^* \)
\lab{ansatz}
\ee 
where the coordinates are such that the metric is 
\be
ds^2 = \( d r\)^2 + \frac{4 r^2}{\( 1+\z2\)^2}\; dz \; dz^*
\lab{metric}
\ee

If one takes $u=z$ and $u^*=z^*$ the decomposition \rf{nicedecomp}
becomes the hedgehog ansatz (\ref{hedge}), with $\zeta\(r\)$ palying
the role of the profile function $f(r)$. So, the skyrmions of unity
charge belong to the integrable sector. The rational map ansatz are
particular cases of the configurations \rf{ansatz}, and so solve the
constraints \rf{staticconst}. However, the rational maps associated to
charge greater than $1$ do not provide solutions for the Skyrme model,
but just approximations to the true solutions. 

Summarizing, we have a submodel of the Skyrme theory with an infinite
number of local conserved currents, and that possesses the charges
$\pm 1$ skyrmions as solutions.

\section{The Skyrme-Faddeev model}
\label{app: scaling}
In view of the above results it is natural to attempt to go from the
widing number charge of $S^3\rightarrow S^3$  to the Hopf map
$S^3\rightarrow S^2$ 
reducing the  target space  to the sphere $S^2\equiv  SU(2)/U(1)$. The
topological charge becomes the linking number of the preimages of
points of $S^2$. This is
what Faddeev proposed,
looking for the string of QCD. The solitons would have then knot
configurations and the simplest
allowed solution would be axially symmetric.
The action for the Skyrme-Faddeev model is then given by
\be
 S = \int d^4x \( m^2  \(\partial {\bf n}\)^2 - {1 \over e^2}\(
\partial_{\mu} {\bf n}
\times \partial_{\nu}{\bf n}\)^2\)
\lab{sflag}
\ee
where  ${\bf n}$ is a $SU(2)$ triplet of scalar field with unit norm,
${\bf n}^2=1$ and $m$ is a parameter with dimensions of mass. 
A potential term can be added \cite {fadedual} to circumvent the
global problems 
with colour in the glueball interpretation. Such  explicit breaking of the
global symmetry was first suggested in
\cite{vW} to avoid spontaneous Goldstone modes, incompatible with the mass
gap of pure QCD. These terms are also required for the pion mass and phenomenological application
in the ordinary Skyrme case.

On the sphere the complex $u$ field of the stereographic projectionit is
very useful
\be
{\bf n} = {1\o {1+\mid u\mid^2}} \, \( u+u^* , -i \( u-u^* \) , \u2 -1 \) \; ;
\qquad \quad
u \equiv u_1 + i u_2 = \frac{n_1 + i n_2}{1 - n_3}
\lab{stereo}
\ee

 The energy for static configurations on
the Skyrme-Faddeev model is easily found
 \cite{Fad,afz},
\be
E= E_1 + E_2
\lab{e1e2}
\ee
with
\br
E_1 &\equiv&  4 m^2 \, \int d^3 x {\mid \nabla u \mid^2 \o { \( 1 + \u2\)^2}}
\nonu\\
E_2 &\equiv& {8\o e^2}\, \int d^3 x {\( \mid \nabla u \mid^4
- \( \nabla u \)^2  \( \nabla u^* \)^2 \) \o {\( 1 + \u2\)^4}}
\lab{e1e2expr}
\er

All models on the sphere $S^2$, independent of the dimension of
space-time, have a convenient
natural formulation of the zero curvature \rf{zccond} in the  approach 
of \cite{afg} given by
\br
A_{\mu} &=& -\pa_{\mu} W \, W^{-1} 
\lab{lacon}\\
&=&
{-i\o{\( 1+\mid u \mid^2\) }} \bigg(  \(\pa_{\mu} u + \pa_{\mu} u^* \)\, T_{1}
+i \(\pa_{\mu} u - \pa_{\mu} u^* \)\, T_{2} +
i\( u \pa_{\mu} u^* - u^* \pa_{\mu} u \) \, T_3 \bigg)
\nonumber
\er
where $W$ is the group element given in \rf{wdef}, $T_i$ being the
usual basis of $SU(2)$, $\sbr{T_i}{T_j}=i\varepsilon_{ijk}\, T_k$.  
To obtain the Skyrme-Faddeev's model equations of motion from
\rf{zccond} one takes 
\be
{\tilde B}_{\mu} = {1 \o{ 1+\mid u \mid^2}}
 \(  L_{\mu}  \, P_{1}^{(1)} -  L_{\mu}^* \, P_{-1}^{(1)} \)
\lab{intpotb}
\ee
with $P_{i}^{(1)}$ being the same as in \rf{bjdef}, and 
\be
L_{\mu} \equiv m^2 \pa_{\mu} u -   
{4 \o e^2}\, {K_{\mu}\o{\( 1+\mid u \mid^2\)^2}} 
\lab{lmu}
\ee 
and $K_{\mu}$ is defined in \rf{kmndef}.

\subsection{The rotor spectrum}

Models on the sphere have also in common an {\it integrable} sector given
by the constraint
\footnote{For the simplest $O(3)$ model in $2+1$ this constraint
generalizes the
Cauchy Riemann conditions of the {\it baby Skyrmion} solution \cite{afg}.}
\be
( \pa u )^2 =0
\lab{const}
\ee
Indeed, if one replaces in \rf{intpotb} $P_{\pm 1}^{(1)}$ by 
$P_{\pm  1}^{(j)}$, with $j$ integer, then the zero curvature
\rf{zccond} gives the 
Skyrme-Faddeev's model equations of motion plus the constraint
\rf{const}. Consequently, such submodel has an infinite number of local
  conserved quantities. 

We observe that the scaling stability of the static solutions under
the Derrick's theorem requires that the tw o terms in the energy in
\rf{e1e2} should be equal 
\be
E_1 = E_2
\lab{stab1}
\ee

For the submodel, the second term of $E_2$ in \rf{e1e2expr} does not
exists and that relation implies
\be
\int d^3x \, \cj = \int d^3x \, \cj^2
\lab{stab2}
\ee
where $\cj$ is
\be
\cj = {2\o {m^2 e^2}}\, {\mid \nabla u \mid^2 \o { \( 1 + \u2\)^2}}
\lab{jdef}
\ee

Therefore, the submodel  presents a rotor like spectrum,
with energy  given by
\br
E = 2 m^4 e^2 \, \int d^3 x \, \cj \( \cj +1 \) 
 = 4 m^4 e^2 \, \int d^3 x \, \cj 
 = 4 m^4 e^2 \, \int d^3 x \, \cj^2
\er

\subsection{Gauge vacua and knots}
\label{app:vBW}
The geometrical formulation of Skyrme-Faddev model contains an intriguing
property, which can be relevant for the connection with the gauge theory,
as observed
recently in the context  of lattice approach \cite {vW} and \cite{BW}.

Writing explicitly the components along the step operators of the auxiliary
flat connection \rf{lacon}
(as in eq. (6.58) of \cite{afg}) as

\be
A_j^1 \equiv {\pa_{j} u + \pa_{j} u^* \o{\( 1+\mid u \mid^2\) }}
\qquad \qquad
A_j^2 \equiv i \; {\pa_{j} u - \pa_{j} u^* \o{\( 1+\mid u \mid^2\) }}
\ee
one has
\br
A_j^1 A_j^1 + A_j^2 A_j^2 = 4 \;
\frac{\mid \nabla u \mid^2}{\( 1+\mid u \mid^2\)^2 }
\er

and
\be
\( A_i^1 A_j^2 - A_j^1 A_i^2\)^2 = 8\;
\frac{\mid \nabla u\mid^4 -
\( \nabla u\)^2\( \nabla u^*\)^2 }{\( 1+\mid u \mid^2\)^4 }
\ee
Consequently, the static energy \rf{e1e2} reads
\br
E&=& \int d^3 x \; \(  \( A_j^1 A_j^1 + A_j^2 A_j^2\) +
 \( A_i^1 A_j^2 - A_j^1 A_i^2\)^2\)\nonu\\
&=&  \; \int d^3 x \; \(
4\, \frac{\mid \nabla u \mid^2}{\( 1+\mid u \mid^2\)^2 } +
8\, \frac{\mid \nabla u\mid^4 -
\( \nabla u\)^2\( \nabla u^*\)^2 }{\( 1+\mid u \mid^2\)^4 }\)
\er
Where $e=1=m$ has been taken (notice that eq. (12) of \cite{BW} corresponds
to $e=\sqrt{2}$)
As observed in \cite{vW} the first term   is formally the functional used
(upon minimization) to fix
non-abelian theories to the so called maximal abelian gauge (MAG)\footnote
{This mimics the abelian Higgs phenomenon  and it should correspond
to the monopole condensation scenario of confinement \cite{tH}}\cite {lat}.This
suggests then that  the minima of the Skyrme-Faddeev,
knot configurations with topological charge given by linking numbers, may
correspond to  the vacua of the
nonabelian theory, fixed to maximal abelian gauge.

Our analysis shows, firstly, that the static energy does not correspond
strictly to the MAG, as it involves
diagonal components from the commutator in the second term. Those
diagonal colour components are  absent 
in the {\it submodel}, since due to the constraint \rf{const}, the
second term involves just $\mid \nabla u\mid^4$, the square of the first term,
which   only has
 transverse colour degrees of freedom. Moreover, it is more simple and it
has a rotor spectrum. Therefore, definite results
about exact (or approximate) solutions of the Skyrme- Faddev model, will be
relevant for the MAG procedure, and {\it vice versa}.

Another result from the analysis is that the commutator term for the
energy of the full model,
involves the diagonal component  as a curl, i.e. as chromomagnetic
potential, since the connection $A$ is flat, which is
relevant for the results of dual variables in the connection with with QCD
\cite{fadedual}.
It also shows that the Skyrme -Faddev energy cannot be  the functional
 given by space integral of $A^2$, which has been  investiagted by numerical
  and analytical methods \cite{a2} and \cite{vW}.
The idea of breaking splicitly the global $SO(3)$ symmetry, has
 been  also discussed
in our approach \cite{Last}. With such an additional potential term, while it
is possible to have infinite conserved currents,
the chances of finding stable solutions are reduced considerably.

\section{Conclusions}
\label{app:con}
We have reviewed  applications of the generalized zero curvature approach,
based on gauge techniques, to the Skyrme theories,
which capture  topological features of the gauge theory. The original
Skyrme theory
\cite{last}, is specially appropriate to understand how the method works and
its difficulties. The results for the {\it integrable sector} of the
Skyrmion Ansatz, found by direct computation in \cite{last}, are
explained and some useful details are provided.

For  the  Skyrme Faddeev model we paid special attention to the
observations that
the auxiliar gauge formalism allows to look at the  Skyrme Faddeev model as
a gauge fixing of the nonabelian theory. Our analysis shows that
the static energy corresponds strictly to the functional minimized in MAG
fixing procedure only in the reduced
submodel, which is more simple and it presents a rotor spectrum. In the
full model it has still diagonal degrees of freedom, of the
chromomagnetic type.
We conclude, in agreement with results from  perturbative
\cite{gies} and lattice 
 methods \cite {vW}, that there
is some evidence for the Skyrme Faddeev model representing global
properties of the pure non-abelian theory in the infrared, but that
 some ingredients are missing
and more work is requiered. And that the generalized zero curvature method
can be useful for that, as it gives physical interpretation
to the gauge dependent quantities from  non-linear models, for  which one
learns in turn from the gauge theory.

\vspace{1cm}

{\bf Acknowledgments.} J.S.G. gratefully acknowledges Fapesp for financial
support, IFT for  hospitality and the organizers for the
stimulating and pleasant atmosphere. L.A.F. is partially supported by
CNPq-Brasil.

\end{document}